\documentclass[twocolumn,aps,prl,showpacs,amsmath,amssymb,floatfix]{revtex4-1}
\usepackage{graphicx}
\usepackage{dcolumn}
\usepackage{bm}
\usepackage{xcolor}

\begin{document}

\title{Pore-size dependence and slow relaxation of hydrogel friction on smooth surfaces}

\author{Nicholas L. Cuccia}
\author{Suraj Pothineni} 
\author{Brady Wu}
\author{Joshua M\'{e}ndez-Harper}
\author{Justin C. Burton}
\email{justin.c.burton@emory.edu}

\affiliation{Department of Physics, Emory University, Atlanta, Georgia 30322, USA}

\date{\today}



\begin{abstract}
Hydrogel consists of a crosslinked polymer matrix imbibed with a solvent such as water at volume fractions that can exceed 90\%. They are important in many scientific and engineering applications due to their tunable physiochemical properties, bio-compatibility, and ultra-low friction. Their multiphase structure leads to a complex interfacial rheology, yet a detailed, microscopic understanding of hydrogel friction is still emerging. Using a custom-built tribometer, here we identify three distinct regimes of frictional behavior for polyacrylic acid (PAA), polyacrylamide (PAAm), and agarose hydrogel spheres on smooth surfaces. We find that at low velocities, friction is controlled by hydrodynamic flow through the porous hydrogel network, and is inversely proportional to the characteristic pore size. At high velocities, a mesoscopic, lubricating liquid film forms between the gel and surface that obeys elastohydrodynamic theory. Between these regimes, the frictional force decreases by an order of magnitude and displays slow relaxation over several minutes. Our results can be interpreted as an interfacial shear thinning of the polymers with an increasing relaxation time due to the confinement of entanglements. This transition can be tuned by varying the solvent salt concentration, solvent viscosity, and sliding geometry at the interface.
\end{abstract}


\maketitle

Hydrogel consists of a solvent-saturated, crosslinked polymer network that exhibits incredibly unique chemical and mechanical qualities. The polymer network allows for a size-selective diffusion of macromolecules, such as DNA, and a macroscopic elasticity coupled with a large maximum strain. By tuning the polymer chemistry, degree of crosslinking, and solvent properties, hydrogels can be used in a wide variety of applications -- for example, agricultural soil enhancement, \cite{Guilherme_2015,Rudzinski_2002,Kim_2010}, medical procedures and replacements  \cite{Greene_2011,Berg_2012,Lieleg_2011,Kisiday_2002,Barth_2016}, biomaterials \cite{Moore_2016,Moore_2015,Moore_2014,Rennie_2005,Lee_2001,Dong_2006,Hamidi_2008,Larson_2016}, and soft robotics \cite{Beebe_2000,Sidorenko_2007,Sun_2012,Bauer_2013,Keplinger_2013,Kim_2015,Yuk_2017}. Many of these applications depend sensitively on the interfacial rheology of hydrogels, where extended polymer chains can interact with both the solvating fluid and solid substrate, leading to a rich spectrum of frictional behaviors \cite{Gong_2006}.  

Generally, the friction coefficients ($\mu$) of hydrogel-hydrogel and hydrogel-solid interfaces are very low, often below $\mu=0.01$ at laboratory-scale sliding velocities. Such low friction coefficients result partially from a hydrogel's high solvent volume fraction and associated lubrication. Yet even without a macroscopic lubrication layer, the polymers in a hydrogel are highly solvated and likely contain a sub-nanometer boundary layer of hydration which allows for molecular slip near the interface \cite{Ma_2015,Raviv_2002}. However, the average ``pore'' size is usually much larger than a molecular hydration layer ($<$ 1 nm), and is of order 5-500 nm \cite{Kim_2010,Johnson_1996,Shoaib_2017}. 

It is important to note that the effective pore size of the network depends on the measurement. We are concerned primarily with fluid flow through the near-surface polymer network. For transport measurements in hydrogels, the ``mesh size'' \cite{degennes} or ``blob size'' \cite{Pincus1976} is often used as the characteristic length scale \cite{Pitenis_2014,Urue_a_2015,Scherer1994,Tsuji2018}. The mesh size, $\xi$, measures the correlation length between monomers in the network, and encompasses thermal fluctuations of the polymer chains. As such, $\xi$ often scales with the average distance between crosslinks in the network. The pore size distribution in hydrogels can also be polydisperse, so that any mesoscale structure in the polymer network will strongly affect its effective hydraulic pore size in response to shear or pressure-driven flow. Near the surface, polymers do not experience an isotropic environment, and can deform due to large shear forces imposed by fluid flow, resulting in non-trivial, macroscale frictional behavior. 

There are three dominant system properties that are known to affect the frictional behavior of a single hydrogel surface contact. First, as mentioned, hydrodynamic forces and polymer-scale deformation will be important. This applies to both a bulk lubricating fluid layer at high velocities, and potential Darcy-like flow through the porous network at lower velocities. For soft materials, the resulting elastic deformation depends strongly on the contact geometry \cite{Saintyves_2016,Pandey_2016,Urzay_2007,Skotheim_2005,Skotheim_2004}. For example, the contact region between a hard sphere and soft solid is curved, and the contact region between a soft sphere and a hard solid is flat \cite{Snoeijer_2013,McGhee_2018}. Second, physiochemical absorption or repulsion between the polymers and the substrate, or between polymers on different hydrogels, will strongly affect the friction \cite{Gong_1998}. Repulsion will tend to draw fluid into the interface, whereas absorption will lead to more solid-like, static friction. Finally, the micro-mechanical and thermodynamic properties of the polymer elastic network, from surface roughness \cite{Yashima_2014} to shear deformation \cite{Kim_2018,Kim_2016,Pitenis_2014,Dunn_2015}, can affect the relaxation and hydration
timescales at the interface \cite{Reale_2017,Moore_2017}.

Here we disentangle many of these competing effects by examining the frictional behavior of common hydrogels, specifically agarose, polyacrylic acid (PAA), and polyacrylimide (PAAm), as a function of sliding velocity. Most of our experiments utilize a hydrogel sphere on a smooth, hard surface, although we also report experiments with a hard sphere on a hydrogel surface, and with two hydrogel surfaces. By varying the normal load, solvent viscosity, salt concentration, and polymer density, we show that friction at low velocities on smooth surfaces is controlled by hydrodynamic shear flow through the porous polymer network, and that the friction coefficient, $\mu$, is inversely proportional to the pore size, $d$. This behavior continues until a critical velocity, $v_\text{c}$, is reached, where $\mu$ precipitously drops by an order of magnitude. Both $v_\text{c}$ and $\mu$ are time-dependent and exhibit nonequilibrium dynamics consistent with a long-time relaxation of the polymer network on the order of minutes, independent of the solvent viscosity. Surprisingly, the friction coefficient can recover in a matter of seconds when sliding is ceased. At higher velocities, $\mu$ is mostly independent of the contact geometry and $d$, as expected in a regime where friction is dominated by a bulk, lubricating layer of fluid. Taken together, these results present a quantitative picture where porous media flow, elastic deformation, and polymer relaxation determine the interfacial rheology of hydrogels under a broad set of conditions.

\begin{figure}[t]
    \centering
    \includegraphics[width=1\linewidth]{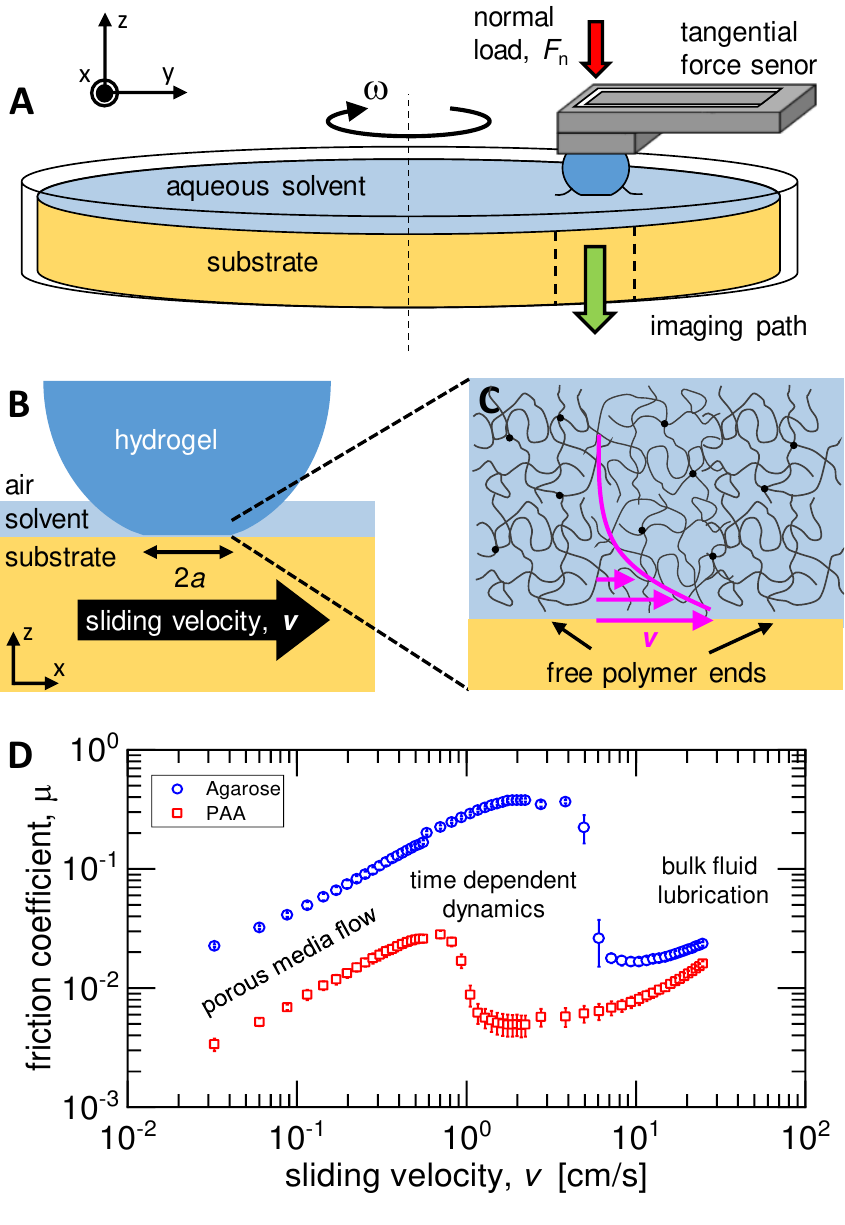}
    \caption{(A) Diagram of the experimental setup showing the spherical sample pressed against the rotating surface. The hydrogel is sheared in the $x$-direction. (B) The characteristic size of the circular contact area is $2a$. (C) The polymer network adjacent to the surface experiences shear as fluid is dragged through it. (D) Plot of $\mu$ vs. $v$ for a 2\% wt agarose gel and a commercial PAA hydrogel particle (JRM Chemical) with $F_\text{n}=0.2$ N and $t_\text{p}$ = 180 s. The substrate was transparent, cast acrylic (poly(methyl methacrylate), PMMA). The three distinct regimes of friction are always observed, although the transitions between them vary with both hydrogel and solvent properties.}
    \label{fig:fig1}
\end{figure}

\section*{Results and Discussion} \label{results}

\subsection*{Frictional regimes on smooth surfaces}

Our experiments used a spherical sample pressed against a hard flat surface with normal force $F_\text{n}$ (Fig.\ \ref{fig:fig1}A, Sec.\ \ref{matmeth}). The frictional force $F_\text{f}$ was localized to a well-defined, circular contact area of radius $a$ (Fig.\ \ref{fig:fig1}B). At each velocity, $\mu$ was measured for an experimental time $t_\text{p}$ before increasing (or decreasing) the velocity. At the lowest velocities, the polymer matrix is adjacent to the surface, separated only by short-range molecular repulsion and any bound hydration layers \cite{Ma_2015,Raviv_2002} (Fig.\ \ref{fig:fig1}C). These hydration layers may affect the local viscosity of the solvent under shear \cite{Laage2017,Halle2003}. Additionally, near the surface, extended polymer chains, some with free ends, are expected to influence the frictional behavior \cite{Dunn_2013,Meier2019}. For hydrogel-hydrogel ``Gemini'' interfaces, these chains may interact and produce a nearly constant or increasing coefficient of friction at low sliding velocities \cite{Shoaib_2017,Pitenis_2014,Dunn_2014}. However, for smooth, hard surfaces, we observe dramatically different behavior, with three distinct frictional regimes, consistent with a few recent studies of hydrogel friction on smooth surfaces \cite{Kim_2016,Kim_2018}.

Figure \ref{fig:fig1}D shows the typical behavior of $\mu$ vs. $v$ for both agarose and PAA. At the lowest velocities, $\mu$ increases smoothly with $v$, implying that $\mu\rightarrow 0$ as $v\rightarrow 0$, which is consistent with a purely hydrodynamic frictional response as the fluid is dragged through the porous polymer matrix. At a critical velocity $v_\text{c}$, the friction coefficient experiences a dramatic drop, often by more than an order of magnitude. Although this behavior bears striking resemblance to the mixed-lubrication regime in a typical Stribeck curve \cite{Lu_2006}, this is perhaps coincidence. In a typical solid-solid contact, as the sliding velocity is increased, a bulk fluid layer begins to penetrate the contact area and drastically reduces $\mu$ through fluid lubrication. In our experiments, however, this sharp decrease in friction appears to be associated with structural changes in the polymer network, which manifests as nonequilibrium, time-dependent behavior in $\mu$. At higher velocities, $\mu$ begins to increase again with $v$, and the properties of the hydrogel polymer matrix and contact geometry are less important. In this regime, our data suggests that our system is well-described by standard elastohydrodynamic friction due to a bulk fluid layer between the hydrogel and the surface. 

\subsection*{Low-velocity regime}

\begin{figure}[t]
    \centering
    \includegraphics[width=1\linewidth]{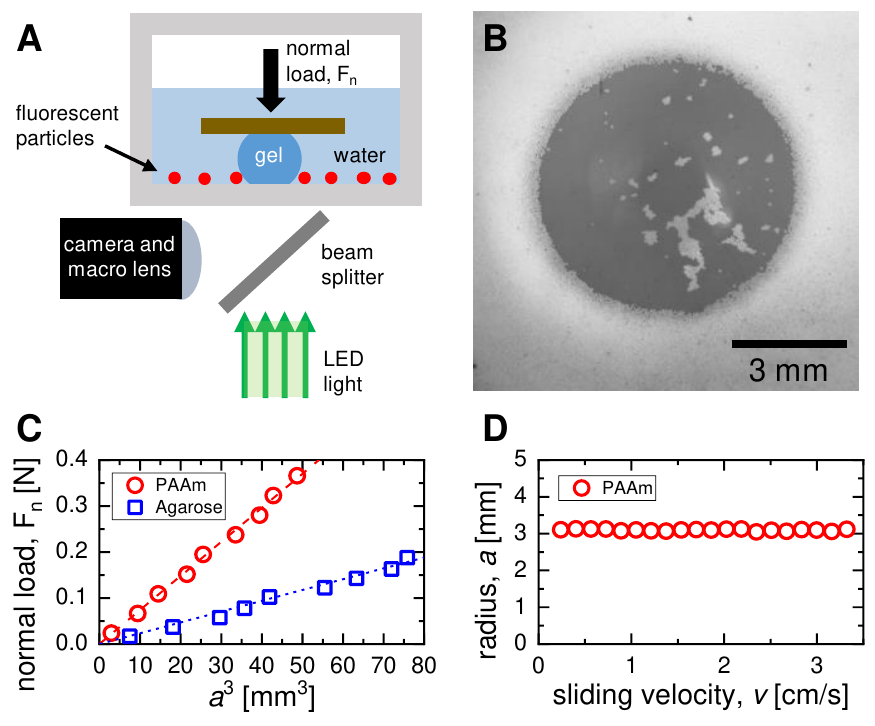}
    \caption{(A) Experimental setup for measurement of the contact area. (B) Red fluorescent particles (mean diameter = 70 $\mu$m) are displaced by the sphere contact. (C) The radius, $a$, of the contact area obeys the Hertzian theory for both agarose and PAAm hydrogels. (D) The contact area was independent of sliding velocity in the experiment.  Error in our measurements is estimated to be $\sim 5\%$ due to the finite size of the fluorescent particles.  Although only data for PAAm is shown, similar results were obtain for agarose hydrogels.}
    \label{fig:fig2}
\end{figure}

At low sliding velocities, we observe a monotonic increase in $\mu$ (Fig.\ \ref{fig:fig1}D). The behavior is consistent with $\mu\propto v^{\gamma}$, where $0.5<\gamma<1.0$. The exponent varied from sample-to-sample, and depended on the type of gel and the polymer concentration, among other properties. We attribute this behavior to the shear force applied to the sphere as fluid is dragged through the porous matrix over a characteristic distance (Fig.\ \ref{fig:fig1}C). The effective shear permeability of the hydrogel surface layers, $k$, may depend on the sliding velocity, and should scale with the effective pore size, $k\sim d^2$. We can estimate the frictional force on the hydrogel using Newton's law of viscosity,
\begin{equation}
F_\text{f}=\eta A \dfrac{v}{d},
\label{fforce}
\end{equation}
where $\eta$ is the dynamic viscosity of the solvent, $A$ is the area of contact, and the velocity gradient has been replaced with $v/d$. We note again that $d$ is the ``hydrodynamic'' pore size at the interface in response to shear, and may vary depending on the heterogeneous structure of the polymer matrix. By measuring $A$ and systematically varying the viscosity, velocity, and polymer concentration, we will show that measurements of $d$ agree well with pore size estimates obtained from transport measurements reported in the literature. 

First, we characterized the circular contact area $A=\pi a^2$ both statically and under sliding contacts. This can be challenging due to the very small optical index difference between the hydrogel and the surrounding solvent. As such, we used the method of particle exclusion microscopy to show the real area of contact between the gel and the acrylic (polymethyl methacrylate) (PMMA) \cite{schulze2016real}.  Figure \ref{fig:fig2}A shows the experimental setup for measurements. We used red fluorescent polyethylene beads with density = 1050 kg/m$^3$ and mean diameter = 70 $\mu$m as indicators of the contact area. When the hydrogel was pressed against the surface, the sedimented beads were displaced and formed a circular ring around the contact area (Fig.\ \ref{fig:fig2}B). Because of the finite size of the particles, the measured contact area was slightly larger than the actual contact area since the particles can only fit within a certain distance from the contact line. Assuming a simple geometry of a sphere penetrating a surface, and given the size of our particles, we estimate that this introduces at most a 5\% error in our contact area measurements. 

\begin{table}[]
\caption{Reduced modulus for the various hydrogels used in our experiments. Concentrations are reported as percent weight. Errors are at most $\pm$5\%, based on the measurement and fitting procedure.}
\centering
\begin{tabular*}{\columnwidth}{l @{\extracolsep{\fill}} cccc}
     \hline
     hydrogel  &E$^{*}$ [kPa] &hydrogel  &E$^{*}$ [kPa]\\ \hline
     0.5\% agarose  & 5  &1.0\% agarose  & 14 \\
     1.5\% agarose  & 26 &2.0\% agarose  & 47	 \\	
     
     8\%, 29:1 PAAm  & 32   & 8\%, 19:1 PAAm  & 26	 \\
     8\%, 9:1 PAAm  & 18 &  12\%, 29:1 PAAm & 41 \\ 
     
     16\%, 29:1 PAAm & 81	& 20\%, 29:1 PAAm & 96 \\	
     24\%, 29:1 PAAm & 155	 & PAA commercial gel & 45\\	\hline
     \label{tab1}
\end{tabular*}
\end{table}

The deformation of the sphere is rather large in most cases. The radius of contact was typically a few millimeters, and the radius of the sphere was 7.5 mm, so $a/R\lesssim0.4$. However, we found that the deformation followed Hertzian contact theory rather well. For a sphere pressed against a flat surface, the Hertz theory predicts that
\begin{equation}
a^3=\dfrac{3 F_\text{n} R}{4 E^*},
\end{equation}
where $E^*$ is the reduced average modulus for the sphere and the substrate:
\begin{equation}
\dfrac{1}{E^*}=\dfrac{1-\nu_\text{g}^2}{E_\text{g}}+\dfrac{1-\nu_\text{s}^2}{E_\text{s}}.
\label{modeq}
\end{equation}
Here $E$ and $\nu$ are the Young's modulus and Poisson's ratio for the gel (g) and substrate (s). Figure \ref{fig:fig2}C shows $F_\text{n}$ vs. $a^3$ for both PAAm and agarose hydrogel samples. The data are well-described by a straight line. Moreover, by measuring $A$ in-situ while the contact is sliding, we found that $A$ remains nearly constant for both low and high sliding velocities (Fig.\ \ref{fig:fig2}D). In most experiments, the Young's modulus of the substrate was much larger than the hydrogel, so that $E^*=E_\text{g}/(1-\nu_\text{g})^2$. The values of $E^*$ for various hydrogels were computed from the fits to the data. The results are shown in Tab.\ \ref{tab1}.

\begin{figure}[t]
    \centering
    \includegraphics[width=1\linewidth]{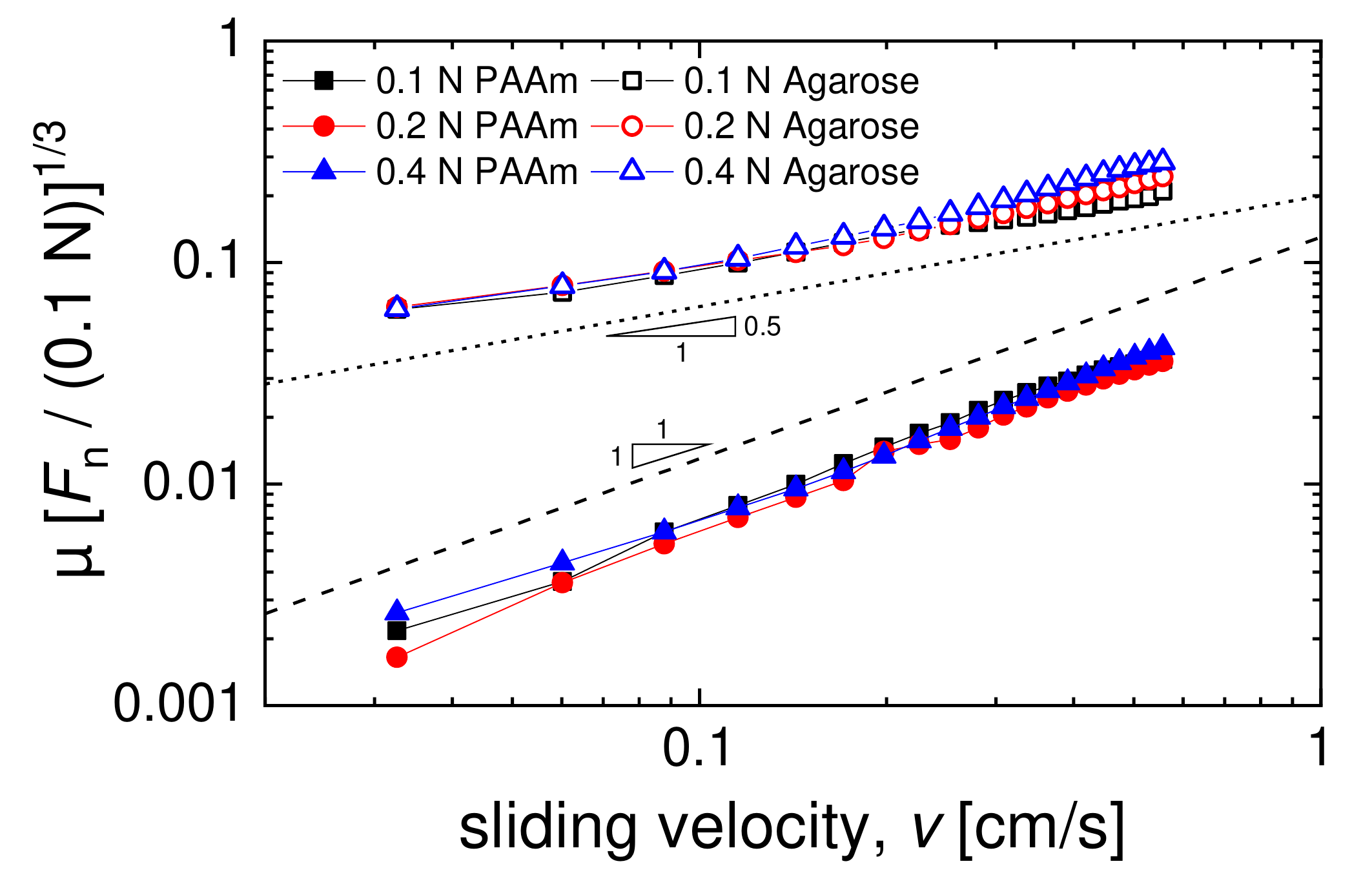}
    \caption{Normalized friction coefficient, $\mu[F_\text{n}/0.1$ N$]^{1/3}$, vs. sliding velocity for both PAAm (12\% wt, 29:1) and agarose (2\% wt) hydrogels at low velocities. The substrate was PMMA, and $t_\text{p}$ = 30 s. The lines show scaling consistent with $\mu\sim v^{1/2}$ (dotted) and $\mu\sim v$ (dashed).  }
    \label{fig:fig3}
\end{figure}

Given the consistency with Hertzian contact theory, for a soft sphere on a hard, smooth surface, we expect that
\begin{equation}
\mu=\dfrac{\pi\eta v}{2d}\left(\dfrac{9R^2}{2(E^*)^2F_\text{n}}\right)^{1/3},
\label{mulow}
\end{equation}
where $d$ can potentially vary with $v$.  We can test some of these dependencies directly. First, the friction should increase as $F_\text{n}^{-1/3}$. This result is quite general and has recently been confirmed for hydrogel-hydrogel interfaces \cite{Urue_a_2018}. Figure \ref{fig:fig3} shows data for both agarose and PAAm hydrogels at 0.1 N, 0.2 N, and 0.4 N. The data is multiplied by $(F_\text{n}/0.1 \text{N})^{1/3}$ in order to collapse the data. Although the range of normal loads is small, the data is consistent with $\mu\propto F_\text{n}^{-1/3}$. The data for PAAm hydrogels is rather linear in velocity, e.g. $\gamma\approx 1$, whereas for agarose, the data is consistent with $\gamma\approx 1/2$. This scaling would imply that for agarose, according to Eq.\ \ref{mulow}, $d\sim v^{1/2}$. This velocity dependence is consistent with previous results showing that the hydraulic permeability ($k$) of bulk agarose increases sharply with velocity due to the strong influence of bound vs. free water in the matrix \cite{Liu_2011}. Similar behavior has been observed for PAAm hydrogels crosslinked with chromium \cite{Grattoni_2001}.

Second, the friction coefficient should increase linearly with the solvent viscosity. Although we tried a number of polar solvents with a higher viscosity than water, we found that solutions of sucrose and water worked the best. In order to confirm that the sugar was penetrating into the gel along with the water, we used small, dry millimeter-sized beads of commercial PAA hydrogels, which took $\sim$ 8 hours to fully swell in water. This process is diffusion-dominated, and thus proportional to the viscosity of the solvent. For example, a particle immersed in an $\eta$ = 15 mPa sucrose solution took $\sim$ 5 days to fully swell. In other polar mixtures, such as glycerol and water, the water preferentially and rapidly diffused into the gel, leaving a higher concentration of glycerol in the solution. 

\begin{figure}[t]
    \centering
    \includegraphics[width=1\linewidth]{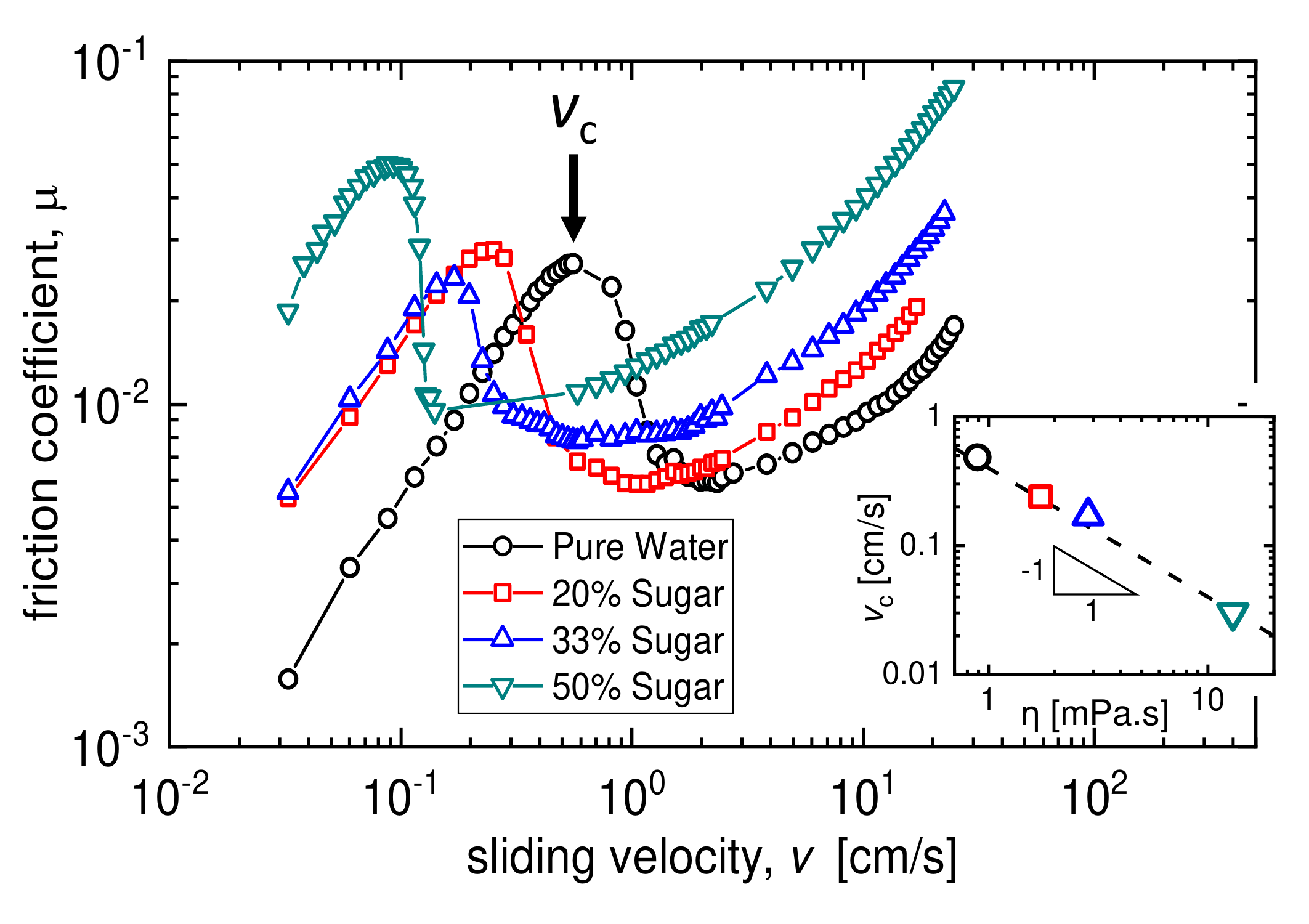}
    \caption{Plot of $\mu$ vs. $v$ for commercial PAA hydrogel particles swollen in sucrose solutions of increasing viscosity. Data is shown for $F_\text{n}$ = 0.2 N on an PMMA surface with $t_\text{p}$ = 180 s.  Error bars are not shown for clarity, but are comparable to those in Fig.\ \ref{fig:fig1}.  Inset: critical velocity $v_\text{c}$ vs. $\eta$. The slope of the dashed line is -1, indicating they are inversely proportional. }
    \label{fig:fig4}
\end{figure}

Figure \ref{fig:fig4} shows $\mu$ vs. $v$ for solvents consisting of pure water and three sucrose solutions. Changing the viscosity of the solvent altered the behavior of $\mu$ in all three regimes. The variation with $\eta$ is consistent with Eq. \ref{mulow} at low velocities, and with elastohydrodynamic theory \cite{Snoeijer_2013} at high velocities, as will be discussed. The dependence on $\eta$ can be seen by plotting the position of the critical velocity at the peak, $v_\text{c}$, as a function of $\eta$, as shown in the inset. For a higher viscosity, the polymers near the surface will experience a higher shear stress, so we may expect $v_\text{c}\propto 1/\eta$ if the transition represents a mechanical stress threshold of deformation for the polymers near the hydrogel-solid interface. 

\begin{figure}[t]
    \centering
    \includegraphics[width=1\linewidth]{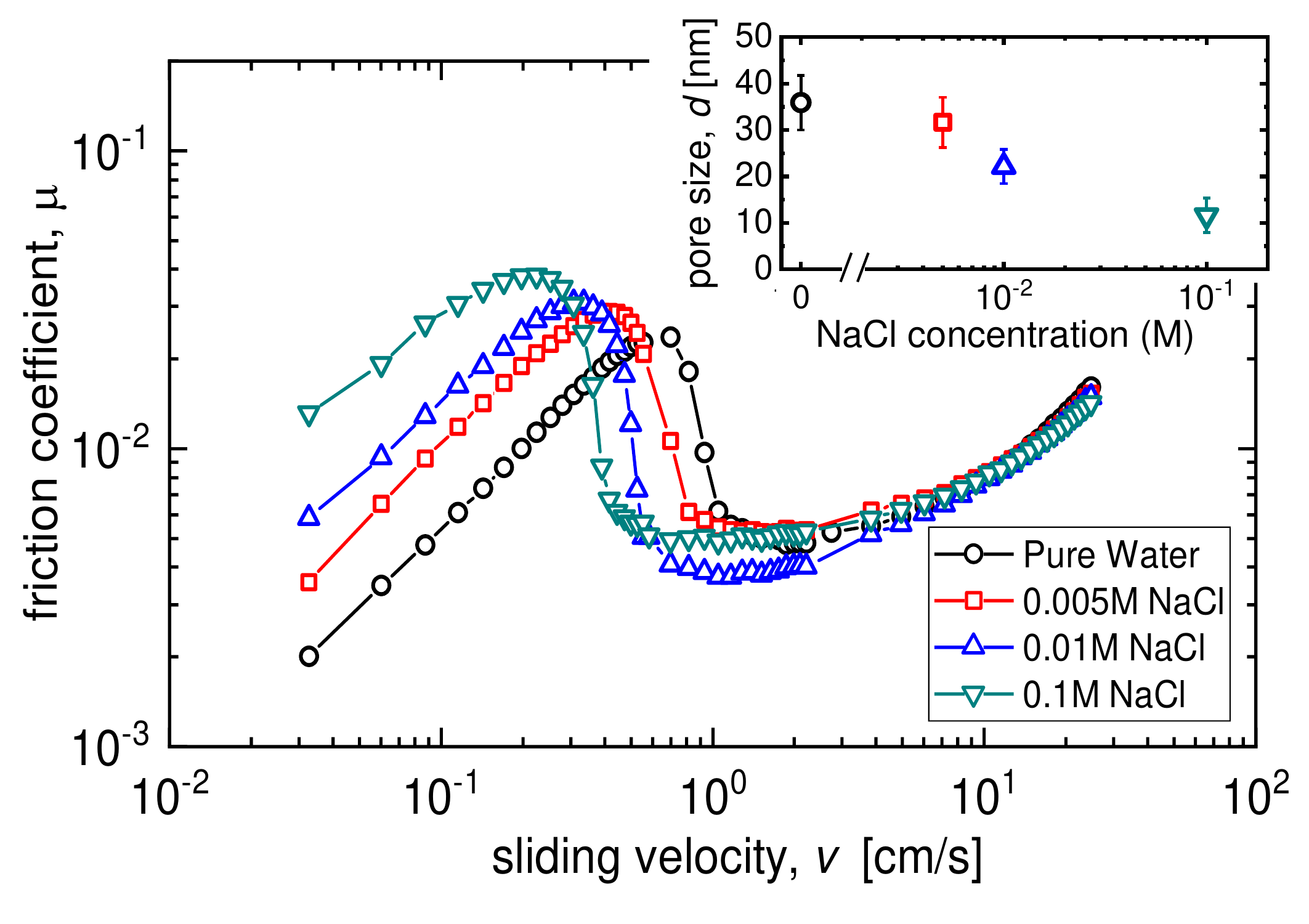}
    \caption{Plot of $\mu$ vs. $v$ for NaCl solutions of increasing concentration. Data is shown for a commercial PAA hydrogel particles with $F_\text{n}$ = 0.2 N on an PMMA surface with $t_\text{p}$ = 180 s.  Error bars are not shown for clarity, but are comparable to those in Fig.\ \ref{fig:fig1}. The inset shows the effective pore size $d$ extracted from the low-velocity regime using Eq.\ \ref{fforce}. The measured moduli for each concentration were $E^*$ = 34 kPa (0.005 M), $E^*$ = 31 kPa (0.01 M), and $E^*$ = 34 kPa (0.1 M).}
    \label{fig:fig5}
\end{figure}

We have plotted the full range of $\mu$ vs. $v$ in Fig.\ \ref{fig:fig4} to illustrate that the data in the high-velocity regime scales more weakly with viscosity than the low-velocity regime. We want to emphasize the fact that the entire range of data \textit{can not} be collapsed onto one universal curve since the mechanisms of friction are distinct in each regime. This is most evident by adding sodium chloride to the water instead of sugar. The addition of salt generally shrinks the polymer matrix and pore size \cite{Horkay_2000,Sivanantham_2012}, so the frictional behavior at low velocities should change. We immersed PAA hydrogel spheres in solutions of NaCl with different concentrations and left them for 5 days so that the salt could fully diffuse into the hydrogel. Figure \ref{fig:fig5} shows $\mu$ vs. $v$ for pure water and three increasing concentrations of NaCl. The friction at low velocities changed dramatically, whereas the friction at high-velocities is basically unaffected. 

From this data and measurements of the contact area $A$ at each salt concentration, we were able to calculate an effective pore size over which the shear flow penetrated into the hydrogel surface. The inset to Fig.\ \ref{fig:fig5} shows that the pore size varied from $\sim 37$ nm for pure water to $\sim 12$ nm for a 0.1 M NaCl concentration. As mentioned previously, for smaller values of $d$, the polymers near the surface experience a higher shear stress, and thus we may expect a decrease in the critical velocity $v_\text{c}$. This is consistent with the data in Fig.\ \ref{fig:fig5} as the salt concentration is increased.

     
     

\begin{figure}[t]
    \centering
    \includegraphics[width=1\linewidth]{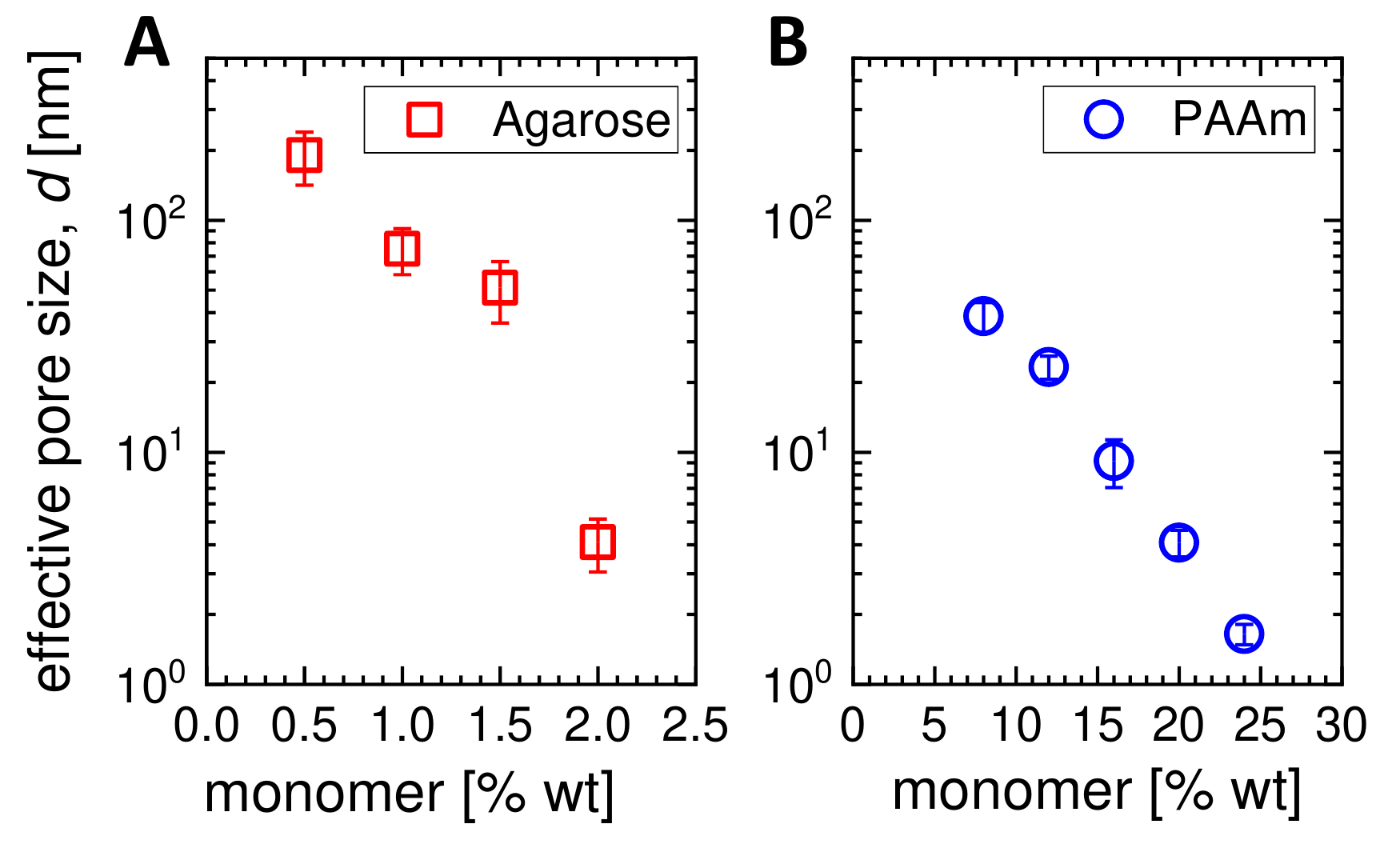}
    \caption{Effective pore size $d$ as determined from Eq.\ \ref{fforce} in the low-velocity regime for $F_\text{n}$ = 0.2 N and $t_\text{p}$ = 30 s. The crosslinker ratio for PAAm hydrogels was held fixed at 29:1. The effective pore size was calculated at each velocity, and then averaged. Error bars represent the standard deviation of the data.}
    \label{fig:fig6}
\end{figure}

We can directly vary the effective pore size, $d$, in our hydrogel samples by varying the polymer concentration during synthesis. The results at low velocities are qualitatively similar to the addition of NaCl. Figure \ref{fig:fig6} shows the computed values of the pore size extracted from data in the low-velocity regime for both agarose (A) and PAAm hydrogels (B). As implied by the sub-linear dependence for agarose shown in Fig.\ \ref{fig:fig3}, the value of $d$ may vary with velocity, so we calculated $d$ at each velocity, and then averaged the data to obtain the mean effective pore size between 0.02 $<v<$ 0.6 cm/s. The standard deviation of this ensemble is represented by the error bars. As expected, $d$ decreased with monomer concentration for both gels, corresponding to smaller pore sizes. We emphasize that this effective pore size is fundamentally a hydrodynamic measurement, and will depend on the connectivity of the pores and their size distribution. This is distinct from estimates of hydrogel pore sizes in the literature that come from the diffusion of macromolecules, such as DNA, during electrophoresis. 

Nevertheless, our measurements of pore size are in reasonable agreement with electophoretic measurements. For example, for the concentrations of both agarose and PAAm used in our experiments, agarose will generally have larger pore sizes, often exceeding 100 nm, whereas 50 nm seems to be typical of PAAm with a $\approx$ 29:1 crosslinker ratio \cite{Stellwagen_1998,Barril_2012,Fatin_2004,Holmes_1991,Narayanan_2006}. A more relevant measurement of pore size is the bulk hydrodynamic permeability using pressure-driven flow. However, the velocities of the fluid in the bulk of the polymer network are much smaller than the sliding velocities used in our experiments, and the permeability of the interfacial layer may be larger than in the bulk fluid layer. 

The hydraulic permeability of agarose hydrogels has been measured directly \cite{Johnson_1996}. For a 1.9\% wt hydrogel, the effective pore size was reported as $\approx$ 25 nm, which is in reasonable agreement with the data for a 2\% wt agarose hydrogel shown in Fig.\ \ref{fig:fig6}A. For PAAm hydrogels, most authors report the permeability $k\sim d^2$. Values in the literature are generally of order $10^{-18}$ m$^2$, corresponding to hydrodynamic pore sizes of order 1 nm \cite{Tokita_1991,Kapur_1996,Grattoni_2001}. We only measure such small permeabilities for our 24\% wt PAAm hydrogel. This is likely due to the difference between bulk and interfacial flow. Fluid must be transported macroscopic distances through the bulk, and is more sensitive to the percolated network of pores, whereas the polymer matrix is less constrained near the interface, resulting in larger pore sizes. 

Finally, we measured $d$ as a function of the crosslinker ratio for PAAm hydrogels. As shown in Fig.\ \ref{fig:fig6}B, for an 8\% wt, 29:1 hydrogel, $d\approx38$ nm. One may expect that increasing the crosslinker concentration would decrease the pore size. But, for an 8\% wt, 19:1 hydrogel, we measured $d\approx14$ nm, and for an 8\% wt, 9:1 hydrogel, we measured $d\approx33$ nm. This non-monotonic behavior was first described by Tokita and Tanaka \cite{Tokita_1991}, and can be explained by an increase in pore heterogeneity. Our gels become somewhat white in color at large crosslinker concentrations. This scattered light indicates large fluctuations in structure of order the wavelength of visible light. Such large pore sizes will dominate the permeability and provide less-resistive paths for fluid transport. 

\subsection*{Transition regime}
\label{transregime}

\begin{figure}[t]
    \centering
    \includegraphics[width=1\linewidth]{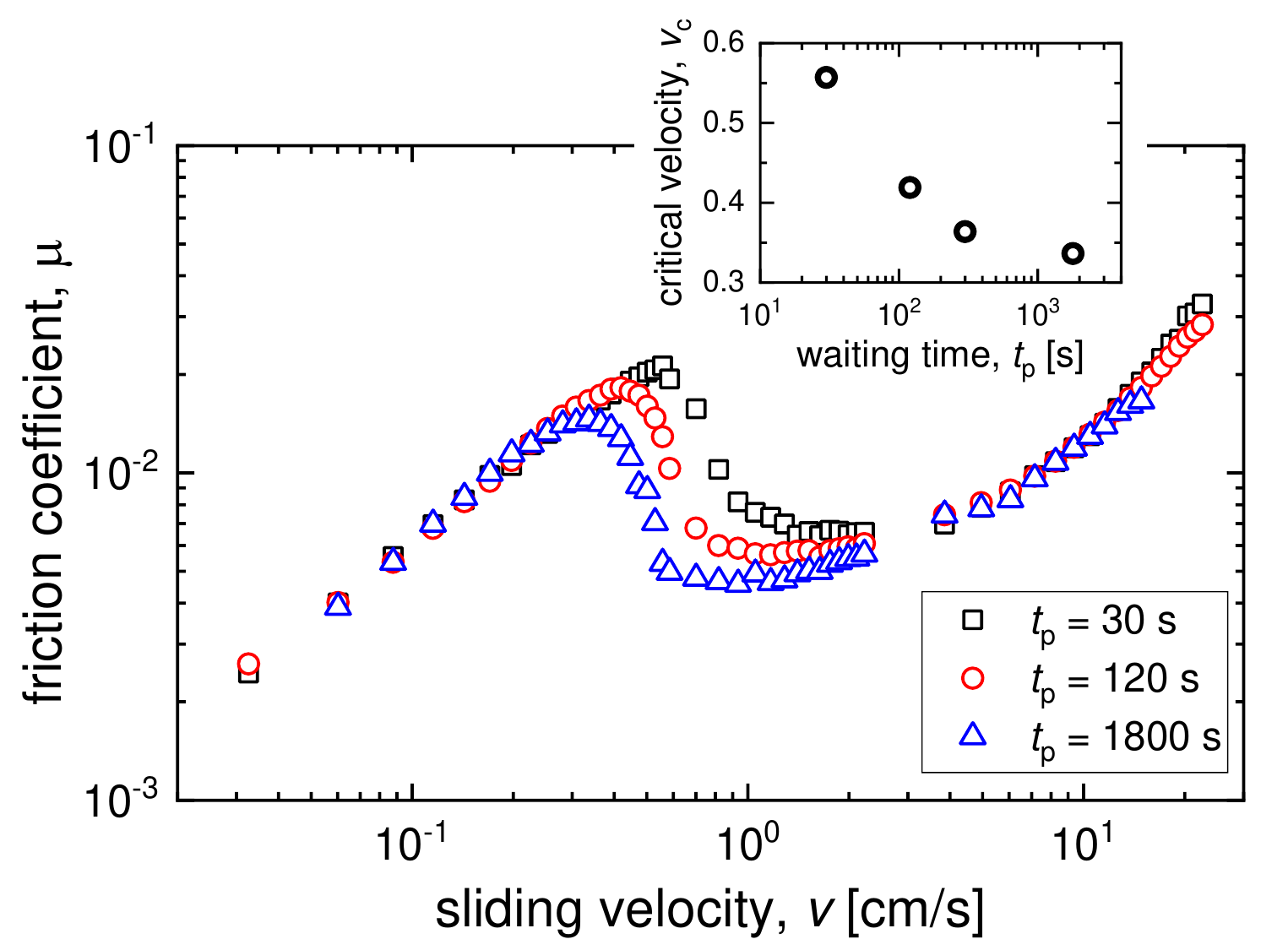}
    \caption{Plot of $\mu$ vs. $v$ for a commercial PAA particle with $F_\text{n}$ = 0.2 N on an PMMA surface for three different values of the experimental running time, $t_\text{p}$, at each data point.  Error bars are not shown for clarity, but are comparable to those in Fig.\ \ref{fig:fig1}.  The inset shows that the value of the critical velocity $v_\text{c}$ decreases  with increasing $t_\text{p}$. The inset contains additional data which is not shown in the main figure.}
    \label{fig:fig7}
\end{figure}

As the sliding velocity and corresponding shear stress are increased, a critical velocity $v_\text{c}$ is reached where the friction begins to decrease rapidly. This velocity represents a threshold stress beyond which structural changes may be induced in the frictional interface \cite{Kim_2016,Kim_2018}. In the traditional Stribeck curve, a sharp drop in $\mu$ corresponds to mixed-boundary lubrication, where a bulk fluid film begins to develop and there is elastic deformation in the materials. The interface deformation must be asymmetric in the sliding direction in order to produce a net lift force from the lubrication layer \cite{Snoeijer_2013,Saintyves_2016,Skotheim_2004,Pandey_2016}. If the transition observed in our experiments was solely due to the formation of a bulk layer, we can estimate its thickness, $h$. For the PAA data shown in Fig.\ \ref{fig:fig1}, $\mu\approx$ 0.005 at $v$ = 2 cm/s. Using a typical contact radius of $a$ = 3 mm, viscosity $\eta$ = 0.001 Pa.s, and $F_\text{n}$ = 0.2 N, we estimate that $h\approx$ 580 nm, which is necessarily larger than the characteristic pore size ($\approx$ 40 nm).

However, the onset and time-dependence of the friction in the transition regime that we observe \emph{can not} be explained by a continuum elastic deformation and formation of a bulk layer, making this frictional transition distinct from traditional mixed lubrication. For example, although the PAA commercial hydrogel and a 2\% agarose hydrogel have similar elastic moduli (Tab.\ \ref{tab1}), the precipitous drop in friction occurs at sliding velocities and frictional forces separated by an order of magnitude (Fig.\ \ref{fig:fig1}). At higher velocities, the curves tend to asymptote to the same values, where the microscopic properties of the porous hydrogel do not matter, but the transition to this regime is characterized by slow relaxation, and is most likely related to the shear-induced deformation dynamics of the polymers near the surface. 

\begin{figure}[h!]
    \centering
    \includegraphics[width=1\linewidth]{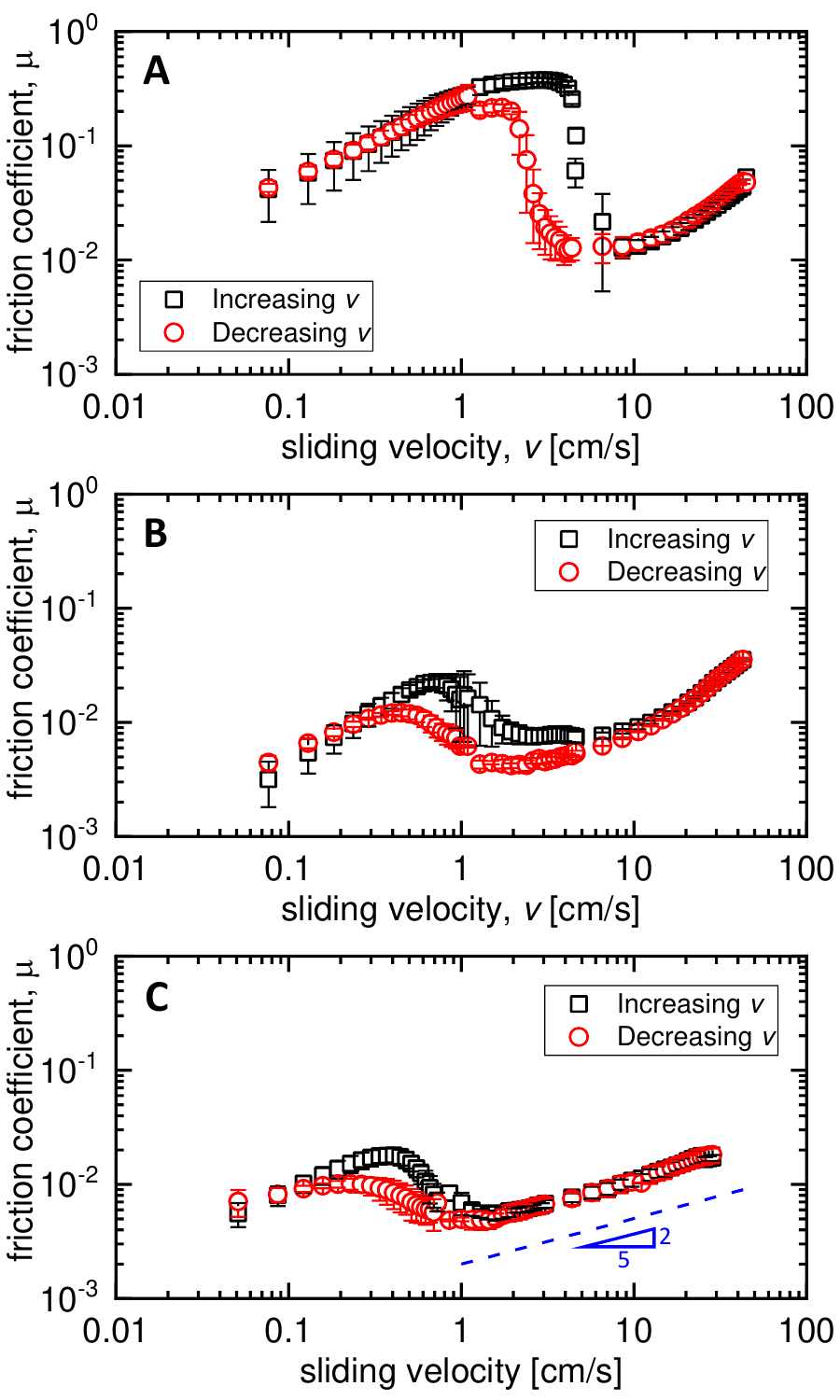}
    \caption{Hysteresis in $\mu$ vs. $v$ for (A) a 1\% agarose particle, (B) a PAA commercial particle, and (C) an 8\%, 29:1 PAAm particle. The experimental parameters were $t_\text{p}$ = 30 s and $F_\text{n}$ = 0.2 N. Black squares show increasing speed, and red circles show subsequent decreasing speeds. The dashed blue line represents a velocity scaling consistent with elastohydrodynamic lubrication theory for soft solids. Error bars here represent the standard deviation of three separate experiments with different hydrogel spheres.}
    \label{fig:fig8}
\end{figure}

\begin{figure}[t]
    \centering
    \includegraphics[width=1\linewidth]{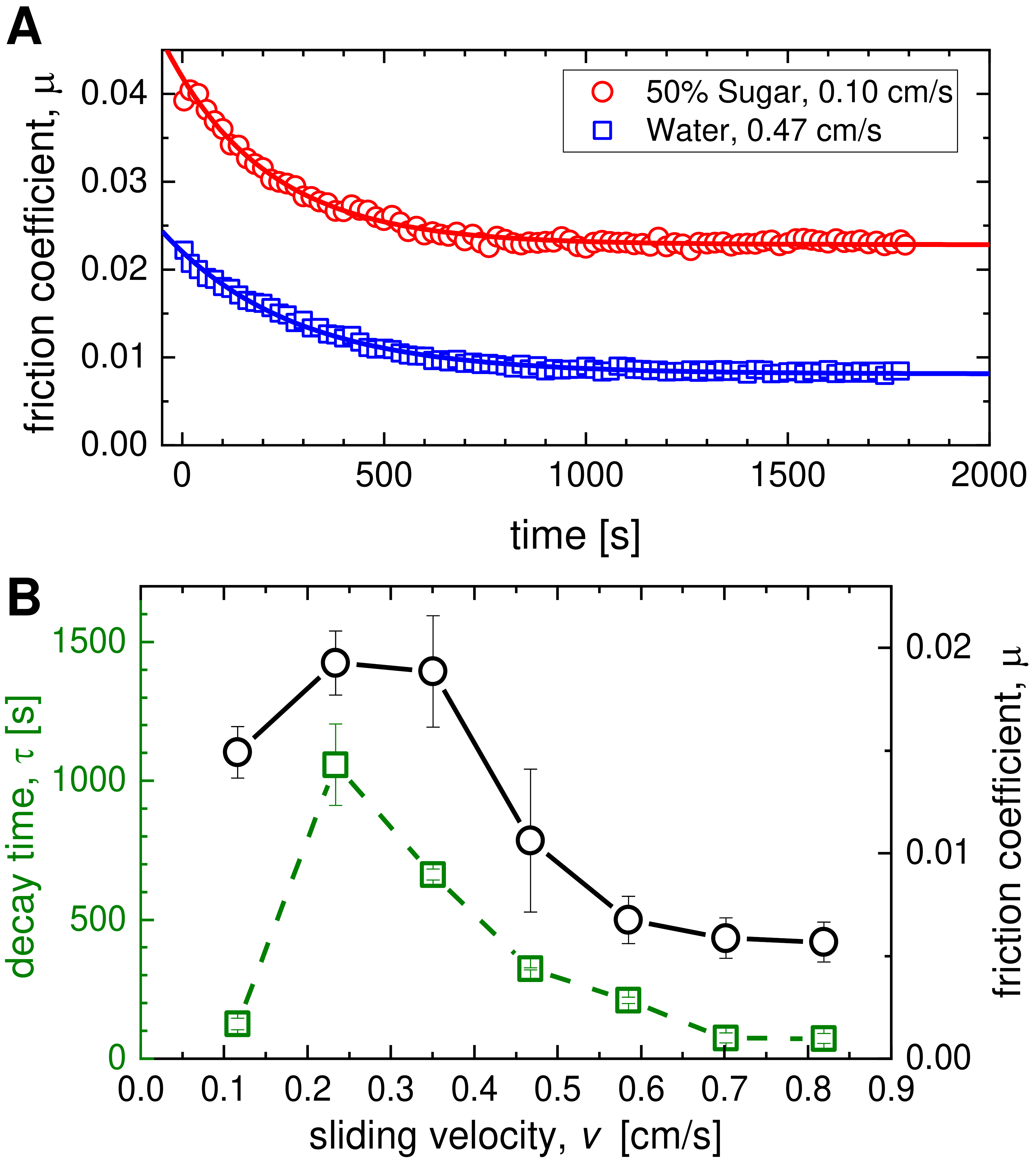}
    \caption{(A) Plot of $\mu$ vs. time for a commercial PAA hydrogel with pure water and one with a 50\% wt sugar solution (the viscosity is $\approx$ 15 times that of water). The solid lines are fits to the form $\mu=(\mu_0-\mu_\infty)e^{-t/\tau}+\mu_\infty$. (B) Decay time $\tau$ and friction coefficient $\mu$ in the transition regime for a commercial PAA hydrogel with $t_\text{p}$ = 1800 s. The data at $v\approx$ 0.47 cm/s is shown in (A). Error bars in $\mu$ represent the variation over 1800 s. Error bars in $\tau$ are from the nonlinear regression performed on the data. }
    \label{fig:fig9}
\end{figure}

This time-dependence is most easily seen in Fig.\ \ref{fig:fig7}, where $\mu$ vs. $v$ is shown for three values of the experimental waiting time $t_\text{p}$. The amount of time spent sliding at each data point doesn't influence the friction at low and high velocities, but it shifts the critical velocity and peak friction coefficient to smaller values. As shown in the inset, $v_\text{c}$ decreases slowly with $t_\text{p}$ before it saturates at large $t_\text{p}$. This variation means that the system is not in equilibrium over experimental timescales, and that the interface is experiencing some form of slow relaxation process over nearly an hour. By increasing and subsequently decreasing the speed in experiments, we also observe a marked hysteresis in the data, as shown in Fig.\ \ref{fig:fig8}. This was true for all hydrogels investigated. A similar hysteresis behavior was observed by Kim et al.\ \cite{Kim_2018,Kim_2016} using an aluminum anulus pressed against a PAAm surface. At both low and high velocities, the data for increasing and decreasing speeds coincided. The amount of hysteresis in the transition regime depended on the time spent at each velocity, $t_\text{p}$, as was illustrated in Fig.\ \ref{fig:fig7}.

For given values of $t_\text{p}$ and $v$, $\mu$ decays monotonically in time, and is often well-described by exponential relaxation, $\mu\approx(\mu_0-\mu_\infty)e^{-t/\tau}+\mu_\infty$. Figure \ref{fig:fig9}A shows data at $t_\text{p}$ = 1800 s for two different hydrogel solvents and velocities. For both cases, the time constant $\tau\approx$ 300 s. Moreover, the relaxation is only visible in the transition regime. Figure \ref{fig:fig9}B shows both the friction coefficient $\mu$ and the decay time $\tau$ in the transition regime. One possible explanation for the appearance of such long timescales in the dynamics of $\mu$ is re-hydration of the lubricating contact \cite{Moore_2017,Reale_2017}. During the low-velocity regime, bulk fluid is excluded from the contact region, and the compressed polymer network is adjacent to the solid surface. As the velocity increases, external fluid begins to intrude into the interface, the polymer network may expand at a rate limited by the imbibition of the solvent. 

We would expect this re-hydration to be driven by a combination of solvent diffusion and mechanical expansion of the polymer network, yet it should be limited by the viscosity of the solvent. However, as Fig.\ \ref{fig:fig9}A illustrates, the time constants for water and an aqueous sucrose solvent that is 15 times more viscous are very similar. In general, we did not observe strong variations in the time constants measured for solvents of different viscosities. Additionally, we found that $\tau$ increased with $t_\text{p}$, meaning that for a longer time spent at each data point, longer relaxation times were more visible in the data. Re-hydration would be driven by a single timescale, whereas the data in Fig.\ \ref{fig:fig9}B suggests that more than one timescale is involved, and the dominant timescale depended on $t_\text{p}$ and $v$. For example, by spending a longer time at each velocity, short-time relaxation processes would equilibriate, and only the long-time relaxation would be visible.

\begin{figure}[t]
    \centering
    \includegraphics[width=1\linewidth]{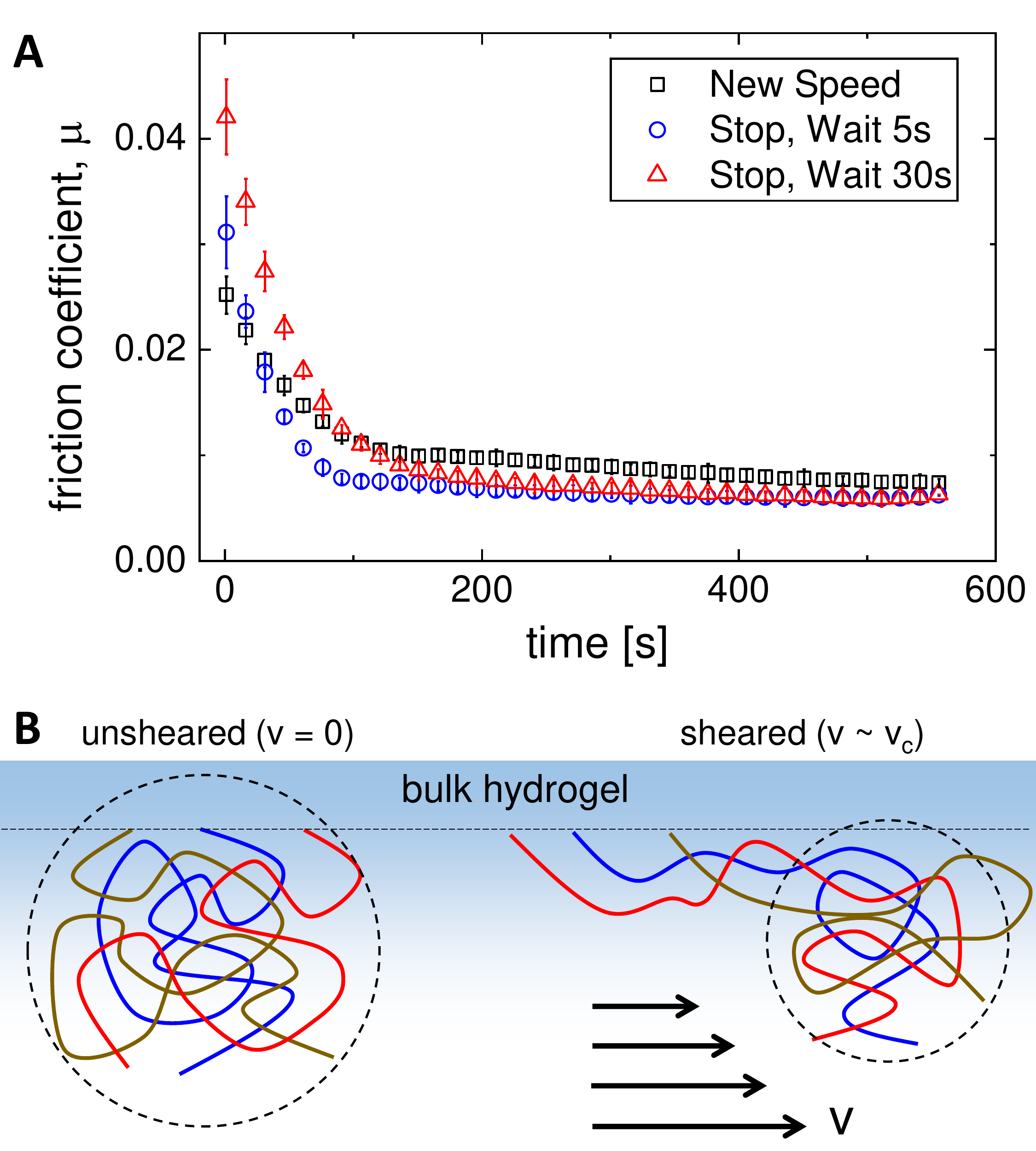}
    \caption{(A) Rapid recovery of friction after quiescence. When increasing to a new speed, after running already at a slower speed, $\mu$ displays a long-time relaxation over hundreds of seconds. Momentarily stopping the experiment ($v$ = 0) for 5 s or more will lead to a recovery of friction, followed by further long-time relaxation. (B) Potential model for frictional relaxation at the interface. Shearing the polymers, as indicated by the direction of the arrow, leads to a confinement of the topological entanglements to a smaller volume, and a dramatic increase in the timescale for polymer dynamics.}
    \label{fig:fig10}
\end{figure}

Strikingly, the relaxation mechanism is highly asymmetric. During an experiment, the interface has experienced successive relaxation events after continually increasing the speed to new values in small steps. At each new speed in the transition regime, we observed a typical long-time decay to some equilibrium value of $\mu$, as shown in Fig.\ \ref{fig:fig10}A. After hundreds of seconds at a given speed, if we then pause the experiment for 5 s ($v$ = 0 m/s) and then continue at the same speed, it seems as if the friction has ``reset'' itself. The friction begins at a larger value and decays again over long times. Increasing this wait time to 30 s leads to an even larger initial friction, and a similar long time decay. This rapid recovery process implies that the mechanism controlling the friction in the intermediate, transition regime is not re-hydration or purely hydrodynamic in nature, as these processes would produce similar timescales for both recovery and relaxation. 

Thus, we propose that the dynamics of the polymer matrix and local, interfacial polymer entanglements play a large role in controlling friction in this regime. A similar conclusion was reached by Kim et al.\ \cite{Kim_2018,Kim_2016} where the authors propose a thixotropic rheological model for the interfacial friction that involves a structural transition in a complex fluid at the interface. The model can well-capture the hysteresis measured in the friction, similar to what we observe in Fig.\ \ref{fig:fig8}. The authors point to polymer chain alignments or interfacial re-hydration as possible mechanisms. Here we build on the former hypothesis since we have shown that re-hydration can not explain our observed relaxation.


Near the surface, well-solvated polymers with free ends are expected to exist. This layer of polymers can extend up to $\sim$ 100 nm from the surface \cite{Dunn_2013}, but this will depend on the hydrophobicity of the surface adjacent to the hydrogel during crosslinking \cite{Meier2019}. In the absence of shear, the polymers are able to explore a wide range of configurations through thermal fluctuations, yet are limited by their local proximity and entanglements. These entanglements are contained within a given volume, as shown by the dashed black line in Fig.\ \ref{fig:fig10}B. Above a critical shear stress, the polymers extend in the direction of the applied stress, reducing the friction in a manner akin to shear thinning. The remaining entanglements are confined to a smaller volume. This confinement can dramatically increase the timescale required to explore configuration space, and thus it will take longer to disentangle further and reduce the friction. A similar increase in the timescales of relaxation occur in molecular glasses and jammed materials due to small increases in density \cite{Liu2010}.  

Although this speculative hypothesis needs to be confirmed with future measurements that can resolve molecular dynamics, it is able to explain the glaring asymmetry in the timescales for relaxation and recovery. When the shear stress is relieved, the polymers retract and are able to quickly explore configuration space again. The balance of thermal fluctuations and shear stress also provides a reasonable length scale for the interfacial dynamics. Assuming a characteristic length scale, $a$, for the polymers at the interface, we can balance entropic stress, $k_\text{B} T/a^3$, with the critical shear stress from the flow, $\eta v_\text{c}/d$ (Eq.\ \ref{fforce}). Equating these stresses, we arrive at:
\begin{equation}
v_\text{c}=\dfrac{d k_\text{B} T}{\eta a^3}.
\label{vceq}
\end{equation}
The characteristic length $a$ is often associated with the mesh size $\xi$ \cite{Pitenis_2014,Urue_a_2015}. Using typical parameters, $\eta$ = 1 mPa.s, $d\approx$ 50 nm, $T$ = 300 K, then length scales of order $a\approx$ 30-50 nm produce critical velocities consistent with our experiments ($v_\text{c}\approx$ 1 cm/s). This is reasonable given the pore size and expected thickness of the surface polymer layer \cite{Dunn_2013}.

This stress balance is somewhat similar to those used in previous studies on hydrogel friction \cite{Kim_2016,Urue_a_2015,McGhee2019}, although we make a distinction between the length scale over which polymers may extend ($a$), and the equilibrium pore size ($d$). Equation \ref{vceq} quantitatively captures the dependence on viscosity and pore size, as shown in Figs. \ref{fig:fig4} and \ref{fig:fig5}. However, the variations in the data with temperature shown in Kim et al.\ \cite{Kim_2016} are not conclusive enough to implicate the role of temperature, and they rightly point out that variations in $v_\text{c}$ due to hysteresis are larger than potential temperature effects. Recently, McGhee et al. showed that temperature changes are consistent with a pore size and viscosity dependence to the friction at low velocities.  Ultimately, a better understanding of the entanglement topology and solvated polymer environment near the surface is needed to provide more quantitative estimation of relaxation timescales in the transition regime.

\subsection*{High-velocity regime}
\label{highvel}

As the velocity is increased well past $\sim$ 1 cm/s, we observe a further increase in friction that is consistent with the hydrodynamics of a bulk fluid layer. The most convincing piece of evidence for this is that nearly all data for a given solvent viscosity and contact area seems to follow the same trend. In Fig.\ \ref{fig:fig1}, despite having a friction coefficient separated by more than an order of magnitude in the time-dependent regime, the data for agarose and commercial PAA hydrogels approach the same values at high velocities. The addition of salt has a large effect on the friction at low velocities, but is unnoticeable at high velocities (Fig.\ \ref{fig:fig5}). However, changing the viscosity of the solvent by adding sugar leads to an increase in friction at high velocities, as expected (Fig.\ \ref{fig:fig4}). 

In this regime, the friction due to hydrodynamic drag should be well-described by elastohydrodynamic lubrication theory (EHL). The soft hydrogel is deformed by the lubrication pressure generated in the thin film, leading to a net lift force that maintains the film thickness \cite{Snoeijer_2013,Saintyves_2016,Skotheim_2004,Pandey_2016}. Under a soft sphere on a hard, flat surface, numerical EHL calculations lead to the following expression for the film thickness $h$ \cite{Hamrock1978,Rennie_2005,Urue_a_2018}:
\begin{equation}
h\simeq h_\text{min}\propto R^{0.77}(\eta v)^{0.65} (E^*)^{-0.44} F_\text{n}^{-0.21},
\end{equation}
where $h_\text{min}$ is the minimum film thickness, and $h$ does not deviate too much from this value throughout the contact area. This scaling agrees well with a recent analytic similarity solution for the two-dimensional flow under a soft cylinder \cite{Snoeijer_2013}:
\begin{equation}
h\simeq h_\text{min}= 0.4467\left(\dfrac{27\pi(\eta v)^3R^4}{(E^*)^2F_\text{n}}\right)^{1/5}.
\end{equation}
For simplicity, we will assume that $h\propto(\eta v)^{3/5}$. Since the shear is occuring in a thin fluid layer, we can assume that $h\sim d$ in Eq.\ \ref{fforce}, and thus find that:
\begin{equation}
\mu\propto(\eta v)^{2/5}.
\end{equation}

Taking all the data together for PAA, PAAm, and agarose, we only find good agreement with this scaling for the softest of hydrogels. For example, Fig.\ \ref{fig:fig8}C shows that a slope of 2/5 is consistent with the data for more than an order of magnitude in velocity. This is for an 8\%, 29:1 PAAm particle, with $E^*\approx$ 32 kPa. For stiffer gels, the scaling with velocity is not well fit by a single power law, and the friction increases more rapidly with velocity. We suspect that for stiffer gels, the asymmetric deformation expected by soft EHL is not as pronounced, meaning that a thinner lubrication layer is required to deform the gel to generate lift, leading to higher frictional force. 

\subsection*{Surface and Geometrical Effects}
\label{surfeff}

\begin{figure}[t]
    \centering
    \includegraphics[width=1\linewidth]{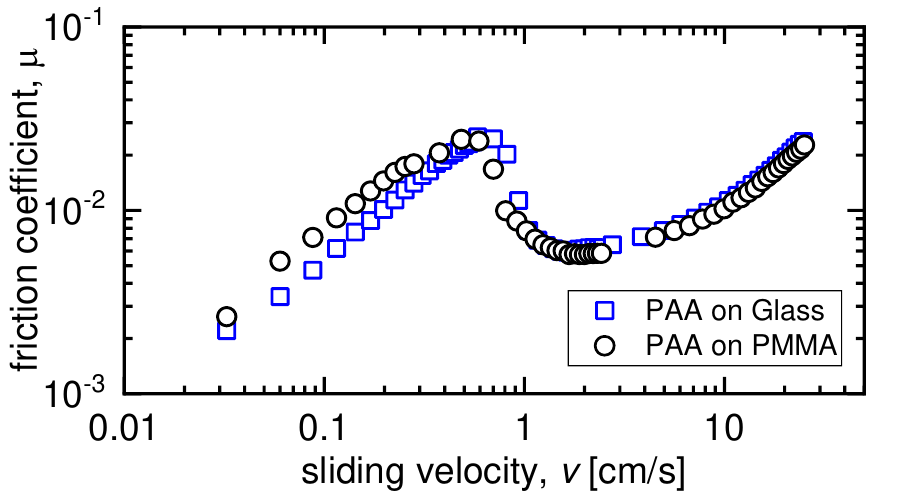}
    \caption{Frictional behavior of commercial PAA hydrogel spheres on both glass and PMMA surfaces.  For all data, $t_\text{p}$ = 180 s and $F_\text{n}$ = 0.2 N.  Error bars are not shown for clarity, but are comparable to those in Fig.\ \ref{fig:fig1}.}
    \label{fig:fig11}
\end{figure}

All of the experimental data shown so far corresponded to various hydrogels on smooth, optically clear PMMA surfaces. However, physiochemical absorption or repulsion between the polymers and the substrate, or between polymers on different hydrogels, can strongly affect the friction \cite{Gong_1998}. We assume that PMMA is fairly hydrophobic, or at least partially wetting, and that there are no specific interactions that would affect the friction. As such, we also tested our gels on smooth glass surfaces, obtaining essentially the same results. Figure \ref{fig:fig11} shows data for a commercial PAA sphere on PMMA, similar to Fig.\ \ref{fig:fig1}, and also for PAA on glass. The data coincides for three orders of magnitude in velocity. Thus, we suggest that our results and analyses will apply to nearly all smooth surfaces which do not have specific polymer-surface interactions. 

In addition, there has been a considerable amount of work done on hydrogel surfaces indented by a frictional probe \cite{Shoaib_2017,Pitenis_2014,Dunn_2014}. These studies show a markedly different behavior at low velocities when compared to our smooth surface, where $\mu\rightarrow 0$ as $v\rightarrow 0$. Pitenis et al.\ \cite{Pitenis_2014} suggested that thermally-driven, transient interactions between polymers at a hydrogel-hydrogel interface can lead to a roughly constant friction at low velocities and Shoaib et al. \cite{Shoaib_2017} suggested that for a smooth probe sliding on a hydrogel surface, friction can increase at low velocites due to a stick-slip mechanism. To test this in our experiments, we fabricated flat, 12\% wt, 29:1 PAAm hydrogel disks as substrates for our spheres. We also fabricated 2\% wt agarose disks. The disks were 2 cm thick, and were cured in a PMMA mold as part of the apparatus. Hertzian contact theory is often employed to describe elastic deformation in both soft and hard surfaces, i.e. it only depends on the composite radius or modulus, as in Eq.\ \ref{modeq}. Thus, we tested the behavior of a hydrogel sphere and an PMMA sphere of the same diameter pressed against both a PAAm and an agarose disk. 

\begin{figure}[!]
    \centering
    \includegraphics[width=1\linewidth]{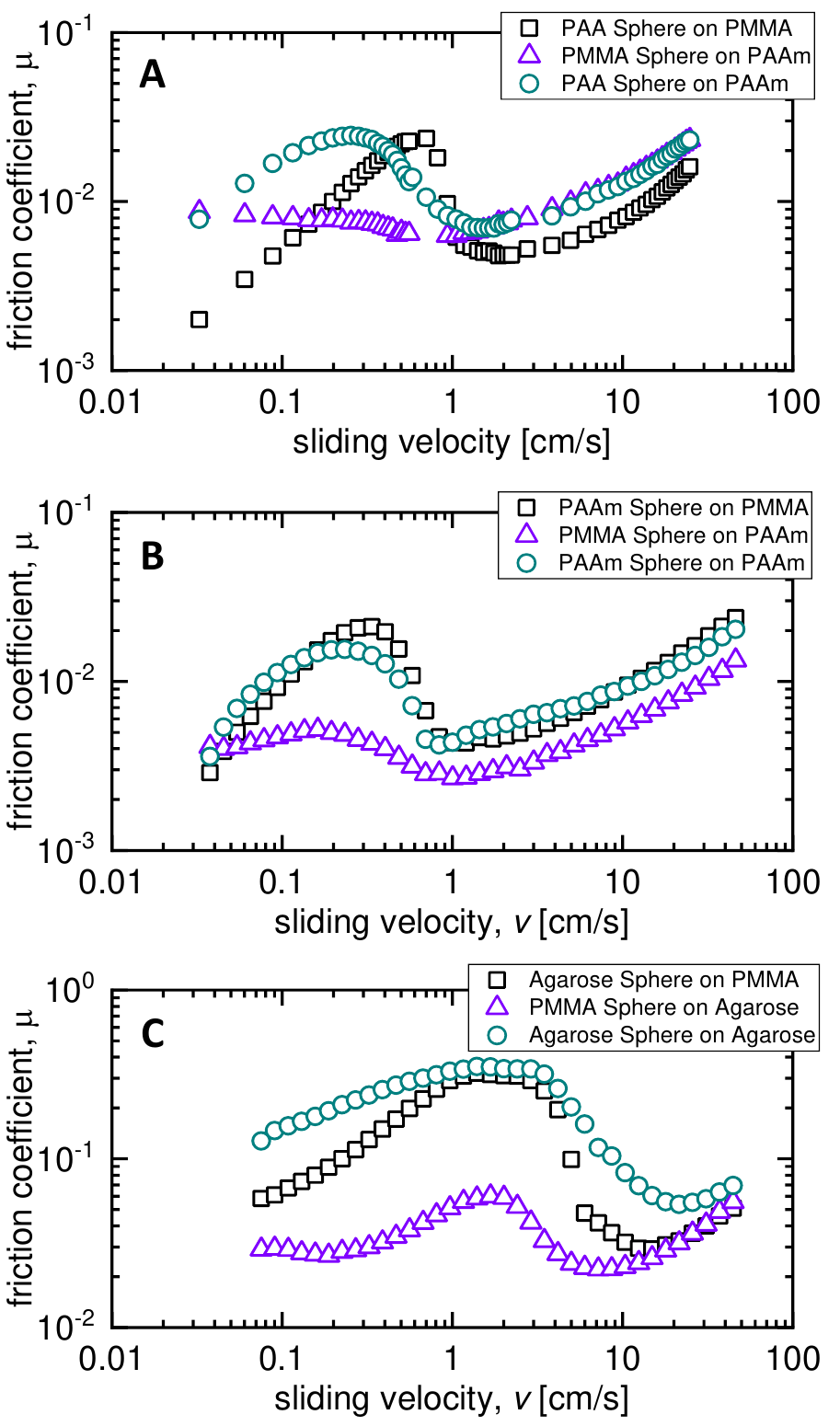}
    \caption{(A) Frictional behavior of a commercial PAA hydrogel sphere on an PMMA disk, a commercial PAA hydrogel on a custom PAAm disk, and an PMMA sphere on a custom PAAm disk. (B) Frictional behavior of a fabricated PAAm hydrogel sphere on an PMMA disk, a PAAm hydrogel on a custom PAAm disk, and an PMMA sphere on a custom PAAm disk. (C) Frictional behavior of a fabricated agarose hydrogel sphere on an PMMA disk, an agarose hydrogel on a custom agarose disk, and an PMMA sphere on a custom agarose disk.  For all data, $t_\text{p}$ = 180 s and $F_\text{n}$ = 0.2 N. Error bars are not shown for clarity, but are comparable to those shown in Fig.\ \ref{fig:fig8}.}
    \label{fig:fig12}
\end{figure}

Figure \ref{fig:fig12}A shows results for a PAA sphere on PMMA, a PAA sphere on a PAAm disk, and an PMMA sphere on a PAAm disk. Since the surface chemisty of the commercial PAA spheres is not the same as our custom PAAm gels, we also tested PAAm spheres on PAAm disks (Fig.\ \ref{fig:fig12}B), and agarose spheres on agarose disks (Fig.\ \ref{fig:fig12}C). At low velocities and intermediate velocities, the three geometries are quantitatively different, and there are two major distinctions to consider. First, the hydrogel disk experiences indentation that must travel with the sliding contact. This can contribute a horizontal force since the frictional surface can be curved around the sphere and is not strictly perpendicular to the sliding direction \cite{Skotheim_2005}, producing higher pressures at the leading edge that affect the gel network \cite{Dunn_2013,Shoaib_2017,Reale_2017}. Second, the polymers near the surface of the hydrogel disk only experience shear transiently as they pass under the contact zone. This means that much of the polymer relaxation processes described previously do not apply. 

Both of these effects can help explain our data.  At intermediate velocities, the friction for the PMMA sphere on PAAm is much lower. This is likely due to a well-hydrated layer on the fresh hydrogel surface that exists prior to its entrance into the contact region. Each part of the hydrogel surface only experiences frictional forces transiently, which is quite distinct from the constant frictional force applied to the contact area under a hydrogel sphere on a smooth surface. At low velocities, $\mu$ tends to increase slightly, and is larger than PAA on smooth surfaces. We attribute this to geometrical effects from the indentation and circular contact area on the PAAm disk. Since hydrogel is not perfectly elastic (fluid must be moved through the network \cite{Reale_2017}), there will be some dissipation inside the gel as it is repeatedly compressed at the leading edge and decompressed at the trailing edge.

For a hydrogel-hydrogel interface (Fig.\ \ref{fig:fig11}C), we observed behavior representative of something in between the flat and curved, indented surface. Since both surfaces are soft, Eq.\ \ref{mulow} would predict a larger $\mu$ at low velocities due to the smaller value of $E^*$, which is consistent with the data. However, the effective value of $d$ should be larger since shear is occurring in both hydrogel networks (sphere and disk), leading to a lower friction. Given the significantly larger values of $\mu$ at low velocities, there must be some geometric effects associated with the indented, hydrogel surface. At intermediate speeds there is still a peak in $\mu$, but it is not as pronounced. At the highest speeds, both indented surfaces associated with the PAAm disk show nearly identical behavior, whereas the smooth surface increases in a similar way, yet with a smaller prefactor. This also suggests that indentation in the soft surfaces leads to a larger overall friction, even at higher velocities. A detailed analysis using soft elastohydrodynamic theory may shed light on this geometric dependence since it is inconsistent with linear elastic theory. 



\section*{Summary and Outlook}

In this study we have investigated the frictional behavior of various hydrogels on smooth surfaces. This was accomplished using a custom pin-on-disk tribometer with optical access to the contact area, capable of measuring friction coefficients below $\mu=0.001$. For all samples, the size of the contact area was reasonably described by Hertz contact theory, and the area did not vary with sliding velocity within our resolution. The friction coefficient followed three general regimes: a low-velocity regime where $\mu$ is determined by the hydrodynamic shear through the porous hydrogel network, an intermediate regime characterized by a sharp drop in $\mu$ accompanied by time-dependent dynamics, and a high-velocity regime consistent with a bulk fluid layer.

At low velocities, the behavior of $\mu$ vs. $v$ provides an estimate of the hydrodynamic pore size in the surface layer. The values computed from our measurements are consistent with known literature values for PAAm and agarose gels measured by both molecular diffusion and bulk fluid flow. At high velocities, some of our softer gels are consistent with predictions from EHL theory, i.e. $\mu\propto v^{2/5}$; however, the friction increases more sharply with velocity for stiffer gels. The friction also depended on the geometry of the interface. A gel ball on a flat disk was significantly different than the inverse, a hard ball on a soft, gel disk. We expect that experiments which are able to optically resolve the nonlinear deformations, as in Saintyves et al.\ \cite{Saintyves_2016}, or full numerical simulations may be able to shed light on this elastic asymmetry and accompanying bulk fluid. 

The intermediate regime, where time-dependent dynamics in $\mu$ are observed, leaves the most open questions. Although the decay in $\mu$ at a specific velocity may be well-described by an exponential behavior, the broad range of relaxation times observed at different velocities is not well-understood.  Although our speculative explanation concerning the available volume for conformational changes, as shown in Fig.\ \ref{fig:fig10}B, is consistent with the long time-scales observed during relaxation and the rapid recovery upon cessation of sliding, it remains to be directly verified. Fluorescent tagging of individual polymers near the surface or other optical techniques which can detect anisotropy of the interfacial polymers in the sliding direction may help resolve these questions. 




\section{Materials and Methods}
\label{matmeth}

\subsection{Tribometer Setup}

Our experiments were performed using the custom pin-on-disk, bi-directional tribometer illustrated in Fig.\ \ref{fig:fig1}A. A spherical sample (PMMA or hydrogel) of radius $R\approx$ 7.5 mm was held stationary to the end of a low-force strain sensor (Strain Measurement Devices S256). This cantilevered spherical sample rested upon a horizontal substrate--PMMA, glass, or hydrogel--to a circular, rotating frame.  The local root-mean-square roughness of the glass and PMMA surfaces were measured to be 2.2 nm and 3.8 nm, respectively, using atomic force microscopy over a 1 $\mu$m$^2$ area. The strain sensor measured the tangential force ($F_\text{f}$) on the sphere, and was calibrated prior to use. A normal force, 5 mN $<F_\text{n}<$ 400 mN, was applied above the point of contact using a fixed amount of mass under the influence of gravity. The friction coefficient was calculated as $\mu=F_{\text{f}}/F_\text{n}$. 

A macroscopic layer of solvent over the substrate (thickness $\approx$ 3 mm) kept the contact hydrated during the experiment. We found that this amount of liquid led to repeatable friction measurements over multiple days, and also minimized any bulk drag from the fluid flow around the sphere. A hole was drilled in the supporting plate below the spinning substrate in order to provide optical access to the frictional contact. The substrate was driven by an geared dc motor, whose gear ratio was changed to achieve a wide range of rotational velocities. Small kinks or gaps in the data (at $v \approx 0.5$ cm/s and $3.0$ cm/s in Fig.\ \ref{fig:fig1}D) are visible when changing gear ratios. The radial position of our spherical sample could also be adjusted to change the velocity range. 

\begin{figure}[t]
    \centering
    \includegraphics[width=1\linewidth]{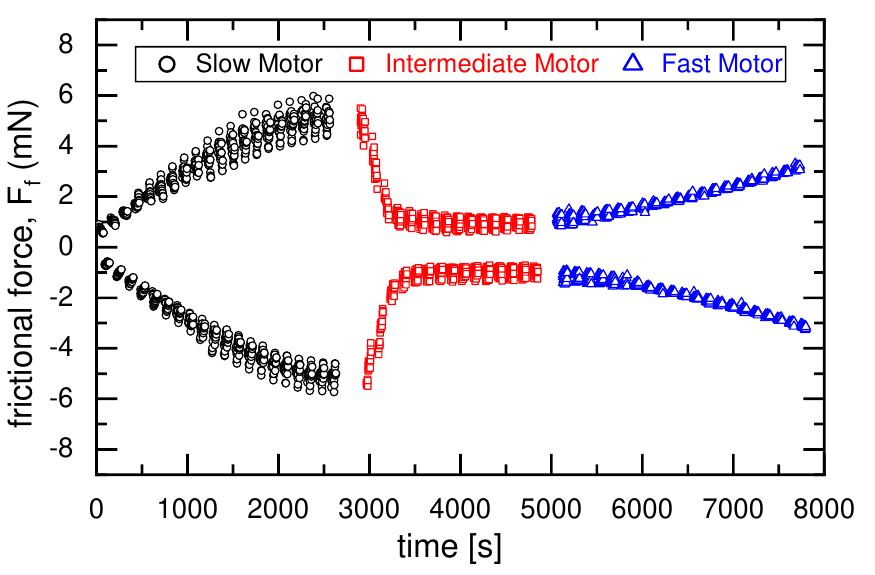}
    \caption{Trace of the frictional force vs. time for a PAA sphere on PMMA disk with $t_\text{p}$ = 60 s and $F_\text{n}$ = 0.2 N. The velocity was continually increased in small increments at time progressed. The stroke lengths for the lowest (0.033 cm/s) and highest (24.8 cm/s) velocities were 1.97 cm, and 1488 cm, respectively. The colors and symbols correspond to data taken with three different gear ratios with different velocity ranges. The larger time gaps in between the black, red, and blue data are associated with the physical time to switch motor gears in the apparatus. Positive and negative data correspond to different directions of angular rotation of the substrate. The noise floor in the force measurement was 0.05 mN. Not all data points are shown for clarity.}
    \label{fig:trace}
\end{figure}

Data acquisition and motor control was accomplished through a custom LabVIEW program. For each data point at a given velocity, we ran the motor clockwise for a period of 30 s $<t_\text{p}<$ 1800 s, paused for 5 s, and then ran the motor counterclockwise for the same time $t_\text{p}$. During each time period $t_\text{p}$, a USB data acquisition device (NI USB-6009) sampled the voltage read-out from the strain sensor at 24 kHz for 0.5 s, and the resulting 12,000 data points were averaged to reduce noise. Thus the friction coefficient was sampled approximately once every 0.5 s. The voltage on the motor (and corresponding velocity at the contact point) was then increased, and the process repeated for the next data point. Measuring the friction in both directions allowed us to calibrate the zero point of the measurement in-situ, and to check for any asymmetries in the frictional measurement in each direction. A typical trace of frictional force vs. time is shown in Fig.\ \ref{fig:trace}.

The data was averaged for each measurement time $t_\text{p}$ in order to provide a single value for the friction coefficient. At some velocities, our data would display an exponential-like decay towards a steady-state value, leading to a large range for that speed's frictional coefficient. We particularly observe these effects within the transition regime when the friction decreased rapidly with sliding velocity.  We also observe that this decaying behavior is far more sensitive to changes in speed than changes in direction. All data were taken at 22 $\pm$ 2$^\circ$C, and reported measurement errors for $\mu$ represent its standard deviation during the experimental time $t_\text{p}$.

\subsection*{Preparation of Hydrogels}

We used two types of hydrogels for our experiments: polyacrylamide (PAAm) and agarose. Although both PAAm and agarose samples were fabricated in the lab, we also used  horticultural hydrogel spherical particles from JRM Chemical (Cleveland, OH).  All chemicals and solvents used for fabrication were purchased from Millipore Sigma. PAAm samples were made by combining the following components, reported as percent weight of the solvent: acrylamide/bis-acrylamide mixture (8-24\% wt), ammonium persulfate initiator (0.15\% wt), and tetramethylethylenediamine (TEMED) catalyst (0.15\% wt). The components were mixed in distilled water and in an oxygen-starved environment for 20 minutes. The solution was poured into custom spherical silicone molds and poly(methyl methacrylate) (PMMA) disc molds for gelation at room temperature over 10 hours. Once gelated, the hydrogel was then immersed in water for 2 hours to remove any lingering chemicals. 

To vary the PAAm cross-linking densities, the acrylamide to bis-acrylamide ratio was switched between 49:1, 29:1, 19:1, and 9:1. To vary the average pore diameters, acrylamide/bis-acrylamide concentrations were made between 8\% wt and 24\% wt. The commercial particles were composed of approximately 70\% polyacrylic acid (PAA) and 30\% acrylamide monomer, as reported by the manufacturer. Spherical agarose samples were made by mixing agarose (0.5-2 g) into 100 ml of distilled water and heating the solution to 60$^\circ$C. The solution was then pipetted into a spherical silicone mold and left aside at room temperature for 30 minutes. The hydrogel was then immersed in water for at least 2 hours. For disks of hydrogel (PAAm or agarose), the gel solution was placed in a PMMA mold of thickness 2 cm and allowed to cure under the same conditions. Finally, we measured the ultimate swelling ratio of the gels after immersing them in water and letting them equilibriate. For example, the commercial PAA particles consisted of 99.2\% water, the 29:1, 12\% wt PAAm particles consisted of 89.4\% water, and the 1\% wt agarose particles consisted of 98.9\% water.

\section{Acknowledgments}
This work was supported by the National Science Foundation through grant NSF DMR-1506446, and also by the Emory SURE program for undergraduate research. This work benefited from fruitful conversations with Alison C.\ Dunn, David Burris, and 
Roger Bonnecaze.

\bibliography{Hydrogel_arxiv}

\end{document}